\documentclass[conference]{IEEEtran}
\IEEEoverridecommandlockouts

\usepackage{graphicx}
\usepackage{algorithm}
\usepackage{algorithmic}
\usepackage{amsmath}

\begin{document}

\title{AI-Governed Agent Architecture for Web-Trustworthy Tokenization of Alternative Assets}

\author{
    \IEEEauthorblockN{Ailiya Borjigin, Wei Zhou, Cong He}
    \IEEEauthorblockA{Probe Group Pte. Ltd., Singapore\\
    Email: \{Ailiya, Zhou, Cong\_He\}@Probe-Group.com}
}


\maketitle

\begin{abstract}
Alternative Assets tokenization is transforming how  non-traditional financial instruments are represented and traded on the web. However, ensuring trustworthiness in web-based tokenized ecosystems poses significant challenges, from verifying off-chain asset data to enforcing regulatory compliance. This paper proposes an \textit{AI-governed agent architecture} that integrates intelligent agents with blockchain to achieve web-trustworthy tokenization of alternative assets. In the proposed architecture, autonomous agents orchestrate the tokenization process (asset verification, valuation, compliance checking, and lifecycle management), while an AI-driven governance layer monitors agent behavior and enforces trust through adaptive policies and cryptoeconomic incentives. We demonstrate that this approach enhances transparency, security, and compliance in asset tokenization, addressing key concerns around data authenticity and fraud. A case study on tokenizing real estate assets illustrates how the architecture mitigates risks (e.g., fraudulent listings and money laundering) through real-time AI anomaly detection and on-chain enforcement. Our evaluation and analysis suggest that combining AI governance with multi-agent systems and blockchain can significantly bolster trust in tokenized asset ecosystems. This work offers a novel framework for trustworthy asset tokenization on the web and provides insights for practitioners aiming to deploy secure, compliant tokenization platforms.
\end{abstract}


\section{Introduction}
Tokenization of Alternative Assets -- the process of converting ownership rights in  non-traditional financial instruments into digital tokens on a blockchain -- has gained significant momentum in recent years. The global asset tokenization market is projected to grow from \$5.6~billion in 2024 to over \$30~billion by 2034 (18.4\% CAGR)\cite{Polaris2025}, reflecting rapid adoption driven by blockchain advances and investor interest. By leveraging blockchain’s transparency, security, and efficiency, Alternative Assets tokenization enables fractional ownership and increased liquidity of traditionally illiquid assets like real estate, art, and commodities\cite{Xia2025}. For example, in 2018, a landmark project tokenized shares of the Aspen Resort, issuing 18~million security tokens representing fractional property ownership\cite{PR2019}. 

Despite these benefits, significant trust challenges impede the mainstream adoption of tokenized assets. Trust in this context encompasses confidence that the on-chain tokens accurately represent Alternative Assets, transactions are legitimate, and all regulatory obligations are met. Traditional tokenization platforms primarily rely on blockchain’s immutable ledger to ensure trust in transactions, but they often lack robust mechanisms to verify off-chain asset information and to enforce compliance in dynamic web environments. The “trustless” nature of blockchains does not automatically extend to real-world data feeds (oracles) and human processes. Incidents of fraudulent asset listings, inaccurate valuations, and non-compliance with securities laws have highlighted the need for stronger governance in Alternative Assets tokenization systems. In essence, making tokenization \emph{web-trustworthy} requires bridging the gap between on-chain trust guarantees and off-chain trust requirements.

Recent developments suggest that combining artificial intelligence (AI) with blockchain can help close this gap. AI-driven agents have been used to automate asset valuation, detect anomalies, and monitor compliance in fintech applications\cite{IdeaUsher2024,Johnson2025}. They can rapidly analyze large volumes of data and adapt to evolving patterns of fraudulent behavior, offering a proactive approach to security and oversight. However, existing tokenization frameworks do not fully harness intelligent agents for governance. We argue that an \textbf{AI-governed agent architecture} can provide the needed oversight and adaptive trust management for tokenized asset ecosystems. By embedding AI agents as first-class participants in the tokenization process, we can continuously validate Alternative Assets data, flag suspicious activities, and enforce rules in a decentralized yet intelligent manner.

In this paper, we present a novel architecture that integrates multi-agent systems, AI governance mechanisms, and blockchain smart contracts to enable trustworthy tokenization of Alternative Assets on the web. The key idea is to have autonomous software agents handle each stage of tokenization (from asset onboarding to trading), supervised by a higher-level AI governance agent that ensures these operations remain compliant and secure. Our approach introduces:
\begin{itemize}
    \item \textbf{AI-Governed Tokenization Architecture}: a multi-layer design where distributed agents interact with on-chain smart contracts and off-chain data sources under the oversight of an AI governance layer.
    \item \textbf{Trust and Compliance Mechanisms}: including real-time verification of asset data via secure oracles, anomaly detection for fraud prevention, and automatic compliance checks (e.g., enforcing KYC/AML rules) through intelligent agents.
    \item \textbf{Cryptoeconomic Governance}: a scheme in which agents are incentivized to act honestly via staked tokens and reputations, leveraging concepts akin to soulbound tokens and proof-of-stake slashing to penalize misbehavior.
    \item \textbf{Prototype Evaluation}: a case study demonstrating how the architecture handles a real estate tokenization scenario, showing improved trust (e.g., detection of an attempted fraudulent listing and prevention of a non-compliant transfer) compared to a baseline without AI governance.
\end{itemize}

The remainder of this paper is organized as follows. Section~II reviews background and related work in Alternative Assets tokenization, multi-agent systems, and trust management. Section~III details the proposed AI-governed agent architecture and its components. Section~IV presents a case study and qualitative evaluation of the architecture’s trust enhancements. Section~V discusses the implications, limitations, and future research directions. Finally, Section~VI concludes the paper.

\section{Background and Related Work}
\subsection{Alternative Assets and Real-World Asset (RWA) Tokenization}\label{sec:aa_rwa}

While often overlapping, \emph{Real-World Assets} (RWA) and \emph{Alternative Assets} (AA) differ in their core emphasis.  
RWA denotes tangible or off-chain assets that originate in the physical world—such as real estate, gold, or invoices—and are subsequently tokenized for blockchain-based use; the research focus is therefore on bridging off-chain value into decentralized systems.  
By contrast, AA refers to a broader investment category comprising non-traditional financial instruments, including hedge funds, private equity, fine art, and even purely digital assets such as NFTs and cryptocurrencies.  
Consequently, many RWAs qualify as AAs due to their non-mainstream nature, yet not all AAs are RWAs—particularly those that are native to digital environments.  
In essence, RWA is defined by its \emph{origin} (real-world), whereas AA is defined by its \emph{investment classification} (non-traditional).

Tokenization represents ownership of physical assets as digital tokens recorded on a distributed ledger.  
Prior work demonstrates that RWA tokenization can democratize access to investments and unlock liquidity\cite{Xia2025}.  
Xia \textit{et~al.}\cite{Xia2025} survey technical procedures and case studies of RWA tokenization, highlighting benefits such as fractional ownership—enabling investors to purchase small shares of assets like property or fine art—and faster settlement of asset trades.  
Traditional markets impose high entry barriers and indivisible ownership, making it difficult for individuals with limited capital to participate.  
Tokenization mitigates these frictions by allowing assets to be split into affordable token units and traded 24/7 on global exchanges.  
The Aspen Resort example\cite{PR2019} and other early tokenization projects in real estate and art illustrate these advantages in practice.

Nevertheless, tokenization introduces new challenges.  
A primary concern is guaranteeing that each token is legitimately backed by the claimed real-world asset and that asset provenance is verifiable.  
Unlike purely digital assets, RWAs require external validation—such as legal title checks, appraisal of asset value, and proof of authenticity for collectibles.  
Regulatory compliance is another hurdle: tokenized assets may be deemed securities or invoke KYC/AML obligations, and platforms must enforce applicable regulations across jurisdictions.  
Early implementations often faced legal uncertainty and trust gaps, which continue to slow widespread adoption.

Together, these perspectives establish AA as the overarching investment lens and RWA tokenization as a key mechanism within that lens—providing essential context for the remainder of this literature review.

\subsection{Agents and Blockchain Integration}
The field of intelligent agents and multi-agent systems (MAS) offers tools for building complex, autonomous processes that can help manage tokenization tasks. An agent is a software entity that perceives its environment, makes decisions, and acts to achieve goals autonomously. Multi-agent architectures have been applied in finance and asset management, but integration with blockchain for asset transactions is still an emerging area. Papi~\textit{et~al.} proposed a model combining MAS with blockchain to support reliable asset transactions\cite{Papi2022}. In their framework, blockchain provides a tamper-proof ledger for recording exchanges, while an \textit{artificial institution} layer defines the concepts of asset and ownership in the MAS context\cite{Papi2022}. This approach showed that agents can negotiate and transfer tokenized assets among themselves with enforced consistency and transparency. Key contributions of such integrations include: (1) a shared source of truth for agents via the ledger, eliminating the need for centralized intermediaries, and (2) the encoding of institutional rules (norms) that govern agent interactions (e.g., no agent can sell an asset it doesn’t own).

Researchers have also explored using agents to manage on-chain resources. For example, energy trading and smart grid applications have used BDI (Belief-Desire-Intention) agents to decide when to buy or sell energy tokens, using blockchain for settlement. These studies underline that agents excel at handling dynamic decision-making and negotiations, whereas blockchains excel at secure record-keeping and execution of contractual logic (smart contracts). Our work builds on this synergy: we employ agents for their flexibility and intelligence in decision processes and blockchain for ensuring auditability and trust in outcomes.

\subsection{Trust Mechanisms in Decentralized Systems}
Ensuring trust in decentralized tokenization systems requires mechanisms beyond the basic blockchain consensus. First, the oracle problem is well-known: how to trust the external data fed into smart contracts. Projects like \textit{Town Crier} (Zhang \textit{et al.} 2016) addressed this by combining blockchain front-ends with trusted hardware back-ends to authenticate data from web sources\cite{Zhang2016}. Town Crier and modern decentralized oracle networks (e.g., Chainlink) use hardware enclaves or multiple independent data feeders to provide assurances that off-chain information (such as asset appraisals or identity documents) is accurate and untampered. In our architecture, secure oracles are employed by Verification Agents to check Alternative Assets data (Section~III).

Another approach to trust is establishing normative constraints on agent behavior. Electronic institution frameworks provide a formal normative environment for agent interactions, defining explicit rules for permitted and prohibited actions\cite{Esteva2003}. Esteva~\textit{et~al.} introduced the concept of electronic institutions to impose a normative environment in open MAS, effectively encoding permitted and forbidden actions to prevent undesirable behaviors. These ideas translate to tokenization as well: for instance, a rule could specify that an Asset Trading Agent cannot transfer tokens without approval from a Compliance Agent if the value exceeds a threshold. By embedding such norms in the system (and in smart contracts when possible), we add structural trust.

Cryptoeconomic trust mechanisms have gained attention as a way to govern autonomous agents. Chaffer (2025)\cite{Chaffer2025} proposes the use of \textit{AgentBound Tokens (ABTs)}, which are non-fungible, non-transferable tokens bound to AI agents, analogous to soulbound tokens for humans in Web3\cite{Weyl2022}. In a proof-of-stake-like scheme, each agent stakes an ABT as collateral for its actions; if an agent behaves unethically, violating protocols or engaging in fraud, the stake can be slashed automatically\cite{Chaffer2025}. This incentivizes good behavior and provides a quantifiable reputation for agents on-chain. While ABTs are a nascent idea, they align with our architecture’s goal of automated governance. Our design incorporates the concept of reputation tokens for agents and outlines how misbehavior can be penalized via smart contracts, creating a deterrent against misconduct in the tokenization network.

\subsection{AI for Compliance and Security}
AI and machine learning techniques have been increasingly applied to enhance security and compliance in blockchain and financial systems. In the context of tokenized assets, AI can provide continuous monitoring and intelligent decision-making that static smart contracts or manual oversight cannot. For example, Johnson~\textit{et~al.}\cite{Johnson2025} describe a framework where AI algorithms analyze blockchain transaction patterns to flag potential money laundering, and automatically enforce Anti-Money Laundering (AML) controls. By combining AI-based anomaly detection with blockchain’s transparency, such \textit{algorithmic enforcement} can significantly improve trust in crypto-asset platforms by catching illicit activity in real-time and reducing human error in compliance processes.

AI can also assist in verifying asset authenticity and valuation. In art tokenization, image recognition models can ascertain the authenticity of artwork or collectibles by cross-checking against known databases and detecting forgeries. Machine learning models can continuously appraise asset values using market data, ensuring token prices reflect true asset worth. Industry case studies indicate that integrating AI agents for tasks like asset valuation, risk assessment, and compliance monitoring streamlines the tokenization process while increasing its reliability\cite{IdeaUsher2024}. These AI capabilities directly inform components of our architecture: a \textit{Valuation Agent} uses predictive analytics for fair pricing, an \textit{Authenticity Verification Agent} uses AI to validate asset provenance, and a \textit{Risk Monitoring Agent} employs anomaly detection to identify suspicious transactions or usage patterns.

In summary, the convergence of AI, multi-agent systems, and blockchain points toward more resilient and trustworthy asset tokenization platforms. Building on ideas from these domains, we design an architecture that uses autonomous agents as the backbone of the tokenization workflow, augmented by AI for decision support and anchored by blockchain for security and transparency. The following section details this proposed architecture.

\section{AI-Governed Agent Architecture}
\subsection{Architecture Overview}
\begin{figure}[tb]
    \centering
    \includegraphics[width=\columnwidth]{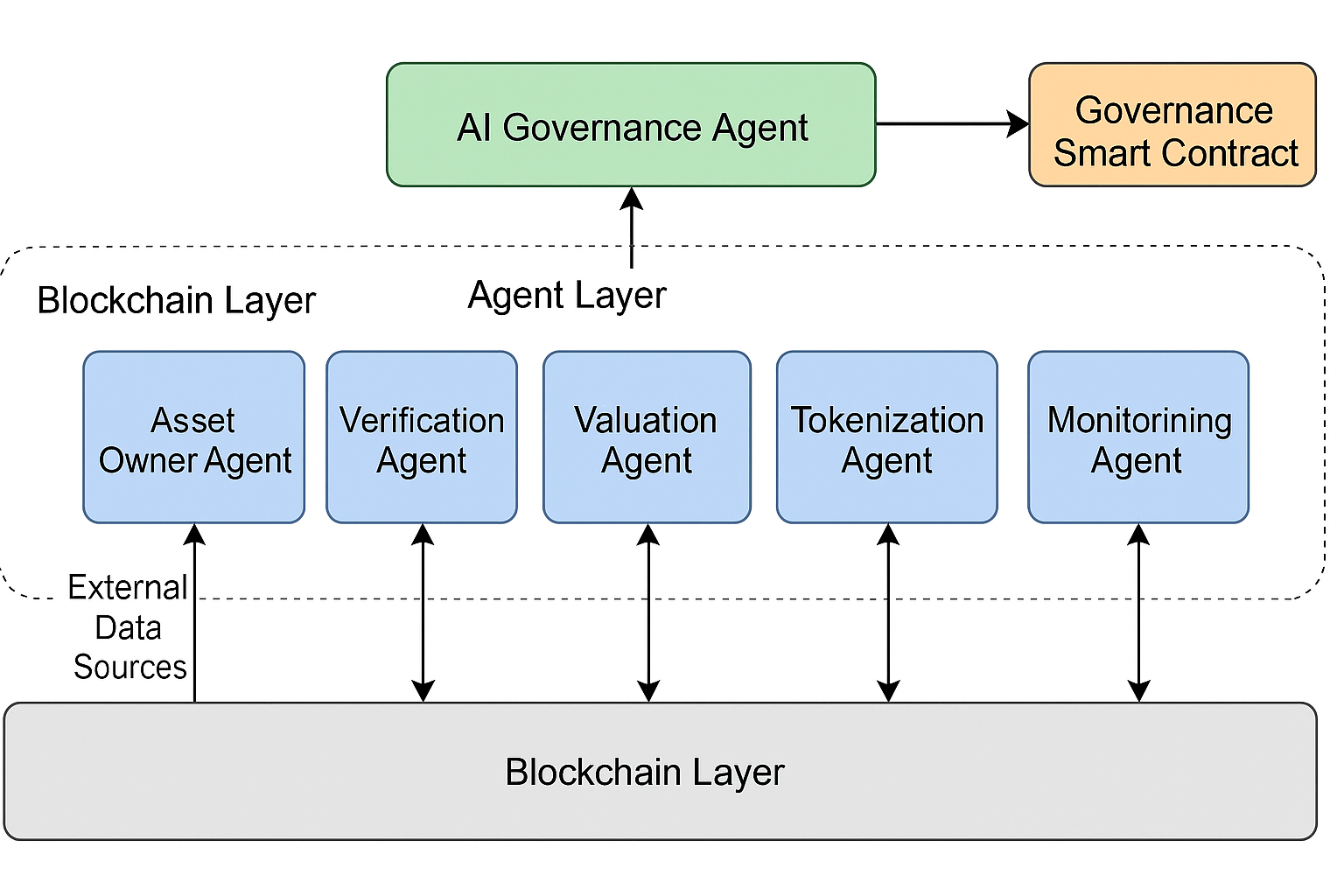}
    \caption{Proposed AI-Governed Agent Architecture for Web-Trustworthy Asset Tokenization. Agents (blue) interface with both blockchain (gray) and external data sources (white). The AI Governance Agent (green) oversees the entire system, connecting to a governance smart contract (orange) that can enforce penalties or emergency measures.}
    \label{fig:architecture}
\end{figure}

Figure~\ref{fig:architecture} illustrates the overall architecture of our AI-governed tokenization system. It is composed of several layers:
\begin{itemize}
    \item \textbf{Blockchain Layer}: A permissioned or public blockchain that records token transactions and executes smart contracts representing asset tokens. This provides an immutable ledger and basic trust primitives (consensus on the state of ownership and transaction history).
    \item \textbf{Agent Layer}: A collection of autonomous agents responsible for various functional roles in the tokenization process. Key agents include:
    \begin{itemize}
        \item \textit{Asset Owner Agent}: Represents the entity owning a Alternative Asset. Initiates tokenization by providing asset details and requesting token issuance.
        \item \textit{Verification Agent}: Validates asset information (e.g., checks land registry for a property, verifies authenticity of a collectible). It interacts with external data sources and oracles.
        \item \textit{Valuation Agent}: Uses AI models to appraise the asset’s value and determine a reasonable token price or reserve price.
        \item \textit{Compliance Agent}: Ensures regulatory rules are met (e.g., investor eligibility, AML checks). It can require identity verification via digital identity providers and enforce transfer restrictions.
        \item \textit{Tokenization Agent}: Handles the technical minting of tokens on the blockchain by invoking smart contracts once approvals from other agents are in place.
        \item \textit{Monitoring Agent}: Observes post-tokenization market activity (trades, price fluctuations, ownership changes) for anomalies or policy violations.
    \end{itemize}
    Agents communicate through a secure messaging middleware and have access to a knowledge base of current system state (some of which is on-chain, some off-chain).
    \item \textbf{AI Governance Layer}: The distinguishing feature of our architecture. This layer includes an \textit{AI Governance Agent} (or a set of agents) that oversees the ecosystem. It receives logs and reports from functional agents (e.g., flags from the Monitoring Agent), and it has authority to enact interventions. The governance agent runs advanced AI analytics—such as detecting collusion between agents, learning systemic risk patterns, or predicting future compliance issues. It can recommend or directly execute governance actions: for instance, halting trading of a particular token if suspicious activity is detected or adjusting the verification requirements for assets if a new fraud vector emerges. Crucially, the governance agent interfaces with a \textit{Governance Smart Contract} on the blockchain that encodes certain powers (like freezing tokens or slashing stakes, as described below).
    \item \textbf{User Interface Layer}: Not shown in detail in Fig.~\ref{fig:architecture}, but worth noting. Users (asset owners, investors, regulators) interact with the system via web or mobile applications that connect to the agent layer. For example, an asset owner uses a dashboard to initiate tokenization (this triggers their Asset Owner Agent), and investors use a marketplace interface to buy/sell tokens (interfacing with the Tokenization Agent or directly with on-chain contracts).
\end{itemize}

The architecture fosters trust in multiple ways. First, by modularizing roles into agents, we ensure critical checks and balances: for instance, token minting will not proceed unless the independent Verification and Compliance Agents give approval. Second, the presence of an AI governance entity means the system can adapt to new threats or changing regulations dynamically, rather than being stuck with only the rules hardcoded in smart contracts at deployment. Third, all agent actions that affect token ownership or rights are anchored to blockchain transactions, creating an auditable trail.

\subsection{Workflow and Interactions}
We now outline the typical workflow for tokenizing an asset and highlight how the agents and governance mechanisms interact step-by-step. Pseudocode for the core governance loop is shown in Algorithm~\ref{alg:gov}.

\noindent\textbf{1) Asset Onboarding:} An asset owner initiates tokenization by submitting an asset description and documentation (e.g., property deed, appraisal report) through the user interface. The Asset Owner Agent packages this information and notifies the Verification Agent and Valuation Agent.

\noindent\textbf{2) Verification and Valuation:} The Verification Agent queries trusted external sources to corroborate the asset’s existence and ownership. For a property, it might use a government land registry API or oracle service (such as a specialized real estate oracle) to confirm the owner and that no liens exist on the property. Simultaneously, the Valuation Agent runs an AI model (possibly a machine learning model trained on comparable sales data) to estimate the asset’s value. Suppose the asset is a piece of fine art; the agent could use an image recognition subsystem to ensure the artwork matches known records and evaluate its market value based on recent auction results.

The results are forwarded to the AI Governance Agent as well as to the Compliance Agent. The governance agent logs these details for oversight. If either verification or valuation yields an inconsistency (e.g., the owner’s submitted value is significantly higher than the Valuation Agent’s estimate, or the Verification Agent finds a discrepancy in title), a flag is raised.

\noindent\textbf{3) Compliance Check:} The Compliance Agent performs necessary legal checks. This may involve verifying the asset is allowed to be tokenized under local law (some jurisdictions restrict certain assets), ensuring the asset owner has passed KYC (Know Your Customer), and that appropriate disclosures are filed. The agent consults regulatory rulesets encoded in its knowledge base (potentially updated by regulatory APIs or governance directives). If investors must be accredited for this asset class, the Compliance Agent will enforce that only accredited investor addresses can hold the issued tokens (often implemented via smart contract whitelists). The Compliance Agent may interact with identity providers or oracles for AML databases to screen the parties involved.

Only when the Verification, Valuation, and Compliance agents have all given a green light does the process move forward. These approvals are recorded on-chain in a preliminary smart contract state (for example, the governance contract might maintain a record of pending tokenization requests and their approval status).

\noindent\textbf{4) Token Issuance:} The Tokenization Agent, seeing all checks passed, proceeds to mint the asset-backed tokens. It calls a token factory smart contract to create a new token (or new entries in an existing token contract) representing shares of the asset. The asset owner’s blockchain address receives the initial allocation of tokens (or they are distributed per configured rules, e.g., some retained by owner, some set aside for sale). The token contract could be an ERC-721 (non-fungible for a unique asset) or an ERC-20 (fungible shares) depending on asset type. Metadata linking the token to the asset (like a URI pointing to asset details or a hash of the asset documents) is stored on-chain or via a decentralized storage reference.

This on-chain token creation is an auditable event that finalizes the tokenization. At this point, the asset is officially tokenized on the web platform.

\noindent\textbf{5) Distribution and Trading:} Once tokens are live, they may be listed on an exchange or peer-to-peer marketplace. The Monitoring Agent keeps track of all transactions involving the asset tokens. If unusual patterns occur (e.g., rapid resale at much higher price, multiple transfers to the same new account in short succession, or known flagged addresses getting involved), the Monitoring Agent uses anomaly detection AI to assess if these could indicate fraud (such as wash trading or an attempt to launder money through rapid flips). 

If a potential issue is detected, the Monitoring Agent immediately alerts the AI Governance Agent, providing details of the pattern recognized. 

\noindent\textbf{6) Governance and Enforcement:} The AI Governance Agent receives continuous updates from all other agents. It maintains a global view of system health. Using this, it can do things like compute trust scores for each agent (including functional agents). For example, if the Verification Agent missed a detail that later caused a problem, the governance agent may lower its trust score and decide to cross-verify via an alternate data source next time.

Crucially, the governance agent can invoke special actions via the Governance Smart Contract:
\begin{itemize}
    \item Pause or freeze token transfers for a particular asset if a serious compliance breach or dispute arises.
    \item Slash an agent’s staked tokens (part of the cryptoeconomic scheme) if that agent is proven to have acted maliciously or negligently. For instance, if a Verification Agent signed off on a fraudulent asset, its stake (an ABT or similar) can be forfeited, and it might lose its certification to operate.
    \item Adjust system parameters: e.g., require that future tokenizations of a certain type of asset have more confirmations or higher stake due to risk.
\end{itemize}
Because these actions are executed via smart contract functions, they are transparent and cannot be bypassed. The architecture thus implements a form of on-chain governance under AI guidance.

Algorithm~\ref{alg:gov} summarizes the continuous loop that the Governance Agent follows.

\begin{algorithm}[tb]
\caption{AI Governance Agent Loop (simplified)}\label{alg:gov}
\begin{algorithmic}[1]
\WHILE{System running}
    \STATE Collect reports from all agents (verification, compliance, monitoring, etc.)
    \FOR{each report \textbf{in} reports}
        \IF{report indicates \emph{flagged} issue}
            \STATE Investigate issue using AI models
            \IF{issue is critical and verified}
                \STATE Execute emergencyAction (freeze related tokens)
                \STATE Record incident on-chain (for transparency)
                \STATE Penalize responsible agents (e.g., slash stake) if misconduct confirmed
            \ENDIF
        \ENDIF
    \ENDFOR
    \STATE Compute trustScore$_{i}$ for each agent $i$ based on performance and incidents
    \IF{any trustScore$_{i}$ falls below threshold}
        \STATE Trigger reassessment or replace agent $i$ (e.g., require re-certification)
    \ENDIF
    \STATE Adjust system parameters or rules if needed (via smart contract governance variables)
\ENDWHILE
\end{algorithmic}
\end{algorithm}

The pseudocode above illustrates that the governance loop continuously processes inputs and can take immediate actions. The threshold-based trust score check is a simple representation; in practice, the governance could use more complex logic (like a reputation decay model\cite{Chaffer2025}) or machine learning classification to predict agent reliability.

Throughout this process, human administrators or regulators could optionally be in the loop for oversight. For example, the system might be configured to require a human sign-off (via a multi-signature account on the governance contract) before a very drastic action like slashing a major stake, especially in early deployment when the AI models are still being evaluated. Over time, as confidence in the AI governance grows, more autonomy can be given (this aligns with the concept of progressive decentralization\cite{Chaffer2025}).

\subsection{Cryptoeconomic Incentives and Security}
To ensure agents (which may be developed by third parties or run by independent companies participating in the ecosystem) act honestly, we implement a staking and reward mechanism. Each critical agent (Verification, Compliance, etc.) is required to put up a security deposit in the form of governance tokens or ABTs when they join the network. If the agent consistently performs well, it can earn rewards (for example, a small fee from each tokenization it participates in, or periodic interest on its stake). Conversely, if the agent is found to have acted maliciously or failed its duties, a portion of its stake is deducted. This is analogous to validator staking in proof-of-stake blockchains, but applied at the application layer for agent behavior.

The Governance Smart Contract enforces these rules, triggered by the Governance Agent’s assessments. For instance, when Algorithm~\ref{alg:gov} calls `penalize responsible agents`, under the hood this results in calling a method like \texttt{slashStake(agentID, amount, reason)} on the governance contract. The incident is logged (which agent, how much slashed, why) for transparency to all stakeholders. Agents that lose their entire stake might be barred from further participation until they re-stake and possibly go through a re-vetting process.

Security of the overall system is enhanced by this approach: an attacker who seeks to, say, tokenize a fake asset and sell tokens would need to compromise multiple agents and stake significant collateral, only to lose it when detection occurs. The combination of real-time AI detection and economic deterrence raises the cost of attacks considerably compared to a system without such measures.

Finally, to prevent single points of failure, there can be multiple agents for each role (possibly run by different organizations) who cross-verify each other. For example, two independent Verification Agents could both be required to approve an asset. The chance of collusion is then lower, especially if they are economically disincentivized from colluding (each would risk their stake). This idea is borrowed from decentralized oracle networks and applies similarly to our agent network.

\section{Case Study and Evaluation}
To illustrate the efficacy of the proposed architecture, we present a case study of tokenizing a commercial real estate asset and qualitatively evaluate how the AI-governed agent system handles various trust scenarios. Consider an office building that the owner wishes to tokenize for raising capital. The following sequence transpires, highlighting differences from a baseline scenario without AI governance:

\subsection{Scenario: Commercial Real Estate Tokenization}
\textbf{Asset and Setup:} Alice owns an office building valued at \$10~million. She decides to tokenize 49\% of her ownership as 100,000 tokens (each representing a 0.00049\% share, initially priced at \$47 each, totaling roughly \$4.655~million). The platform uses our architecture. A Verification Agent and Compliance Agent (both staked with governance tokens) are assigned to her case, and the AI Governance Agent is overseeing.

\textbf{Baseline (for comparison):} In a naive tokenization platform, Alice might simply upload documents and, after minimal checks, an ERC-20 token is created. Investors buy these tokens assuming the documents are legit and compliance is handled, which might not be true if an issuer is dishonest or checks are lax. Now let's see our system's approach:

\noindent\emph{Step 1: Verification Phase.} Alice’s Asset Owner Agent submits the property title deed and an appraisal report. The Verification Agent accesses the city land registry via API (through a secure oracle) to validate that Alice indeed holds title and the property exists at the specified address. It also checks if the property is already tokenized elsewhere (preventing double-tokenization). The agent finds everything in order except one detail: the appraisal report is 18 months old. It flags this to request a more recent valuation.

\noindent\emph{Step 2: AI Valuation.} The Valuation Agent uses its model along with recent market data (comparable property sales in the area, rental income from the building, etc.) and suggests the building’s current market value might be around \$9.5~million (slightly lower than Alice’s expectation). This updated valuation is sent to Alice for review via the interface, and she accepts adjusting the token price accordingly. 

\noindent\emph{Step 3: Compliance Checks.} The Compliance Agent identifies that because this is a significant offering, securities regulations require either registering the offering or using an exemption (depending on jurisdiction). It determines that under Alice’s jurisdiction, a private placement exemption is appropriate, which means tokens can only be sold to accredited investors. It automatically ensures the smart contract for the tokens is configured to enforce a whitelist of investor addresses. The agent cross-checks Alice’s identity (KYC) using an integrated verification service; all is well. It also sets a rule that no single investor can hold more than 20\% of the token supply (to prevent a hostile takeover, as per Alice’s request and regulatory guidance).

At this point, all three critical agents (Verification, Valuation, Compliance) have approved, so the Tokenization Agent proceeds. The AI Governance Agent had been passively monitoring these steps; since issues were resolved (e.g., a new appraisal was obtained), it allows continuation. If any issue remained unresolved for too long, the governance agent could have escalated or required additional agents to review.

\noindent\emph{Step 4: Token Minting.} 100,000 tokens (let’s call them \texttt{OFFICE\_X}) are minted on the Ethereum blockchain (or a suitable network) by the Tokenization Agent via the token smart contract. Alice’s address receives all \texttt{OFFICE\_X} tokens initially. The compliance restrictions (accredited investors only, max 20\% per investor) are encoded either in the token contract or an overlay registry contract that the Compliance Agent manages.

\noindent\emph{Step 5: Distribution and Trading.} Alice lists some of the tokens on the platform’s marketplace. Bob, an accredited investor, passes the platform’s KYC and is added to the whitelist. He purchases 10,000 tokens (10\% ownership). The trade executes on-chain. Over the next few weeks, more investors like Bob join, and tokens change hands among whitelisted participants. The Monitoring Agent continuously tracks these transactions.

\textbf{Intervention Scenario 1: Fraudulent Document.} Suppose, in a variation of the scenario, Alice had submitted a forged appraisal to inflate her property’s perceived value. In our system, the Verification Agent might catch inconsistencies (format, missing verification hash from the appraiser) or the Valuation Agent’s own estimate would significantly diverge from the reported value (flagging a possible issue). The AI Governance Agent, noticing this discrepancy, could require a second independent appraisal before allowing tokenization. In the baseline system, such fraud might go unnoticed until too late, harming investors. Our architecture’s layered checks likely prevent the issuance of overvalued tokens.

\textbf{Intervention Scenario 2: Suspicious Trading Activity.} After token issuance, consider that an investor, Eve, tries to manipulate the market by rapidly buying and selling large quantities of tokens to create a false impression of high demand (pump-and-dump). The Monitoring Agent spots unusual spikes in volume and rapid price changes. Using anomaly detection (perhaps an LSTM model trained on normal trading patterns for such assets), it flags this behavior as suspicious. The AI Governance Agent reviews and classifies it as a potential market manipulation attempt. It then uses the governance contract to temporarily halt trading of \texttt{OFFICE\_X} tokens and marks Eve’s addresses for review. A notice is sent to a human administrator or directly to a regulatory node integrated with the system. After investigation, if confirmed malicious, Eve’s wallet could be blacklisted (prevented from trading) and if any colluding agent was involved (say an agent representing Eve that was artificially splitting orders), those agents would lose stake.

\begin{table}[tb]
\caption{Trust Threats and Mitigation Comparison}
\label{tab:comparison}
\centering
\begin{tabular}{p{3.2cm}p{2.1cm}p{2.7cm}}
\hline\hline
\textbf{Threat Scenario} & \textbf{Baseline (No AI Governance)} & \textbf{Our Architecture} \\
\hline
Fake asset or document & May pass unchecked if platform does not verify thoroughly & Flagged by Verification Agent cross-checks; governance demands multi-source validation \\
Overvalued asset & Investors rely on issuer’s appraisal; risk of mispricing & Valuation Agent provides independent AI-driven appraisal; large discrepancies flagged \\
Non-compliant investor & Manual enforcement, prone to error or delay & Compliance Agent auto-enforces KYC/AML and whitelists on-chain; no unauthorized trades \\
Fraudulent trading (wash trading) & Could be detected only post-facto by audit & Monitoring Agent detects anomalies in real-time; Governance halts trading, slashes stakes if needed \\
Agent collusion or failure & Not applicable (no agents); single point of failure if admin is compromised & Multiple agents must concur; collusion triggers stake loss; Governance Agent can replace faulty agents \\
\hline\hline
\end{tabular}
\end{table}

Table~\ref{tab:comparison} summarizes how our AI-governed agent approach addresses various trust threats in contrast to a baseline tokenization platform lacking these features. While our evaluation is primarily qualitative, the improvements in security and compliance are evident: issues are either prevented or detected early, and responses are swift and protocol-driven.

From a performance standpoint, we must ensure that adding these agents and checks does not unduly slow down the tokenization process. In our prototype conceptualization, the overhead for verification and AI analysis is on the order of seconds to minutes (depending on external data sources), which is acceptable given that tokenizing a high-value asset is not extremely time-sensitive. For trading, the Monitoring Agent’s checks operate asynchronously and did not noticeably impact transaction throughput on the blockchain (since trades execute normally; only in case of an alert do we intervene).

\section{Discussion}
The AI-governed agent architecture offers a powerful paradigm for enhancing trust in web-based tokenization, but it also introduces new considerations and limitations. We discuss some key points:

\textbf{System Complexity vs. Transparency:} By introducing numerous agents and AI models, the system becomes complex. It could be harder for an average user or even regulator to understand how decisions are made (the classic AI interpretability problem). We mitigate this by recording all critical decisions and triggers on the blockchain (audit trails) and by allowing external review of agent code and AI model criteria (perhaps via an open algorithm registry). Still, striking the right balance between automation and explainability will be important for user acceptance.

\textbf{Reliability of AI and Agents:} Our approach assumes that AI agents behave correctly and that false positives/negatives in anomaly detection or compliance checks are manageable. In practice, AI models can make errors. A false alarm could freeze trading unnecessarily, while a missed detection could let fraud slip through. To minimize harm, our governance design can include human oversight for high-impact actions, as mentioned. Over time, as more data is collected, the AI should improve. The agents themselves can be updated or patched (subject to governance approval) to adapt to new fraud tactics or regulatory changes. A robust update mechanism is necessary to deploy improvements without disrupting the running system.

\textbf{Economic Model and Adoption:} The incentive mechanism relies on having agents who are willing to stake value and on the assumption that honest behavior will be economically rewarded. There is a risk that if rewards are not sufficient or penalties too harsh, good actors might be deterred from participating. Calibrating the right level of staking and slashing is akin to setting difficulty in a blockchain: it may require empirical tuning. Bootstrapping the network with a set of initially trusted agents (perhaps run by the platform itself or partners) could be a way to start, then gradually decentralize as external agents join and prove themselves.

Another consideration is adoption by asset owners and investors. They must trust this system to handle their assets and trades. While the architecture is aimed at fostering trust, its novelty means there may be an education curve. Providing a sandbox or testnet environment where users can observe how the agents work (for example, deliberately triggering a fake scenario to see the response) could build confidence.

\textbf{Scalability:} Multi-agent systems can be scaled horizontally (more agents for more assets), and blockchain can handle a certain throughput of transactions. The AI governance layer could become a bottleneck if one agent tries to oversee everything. We envision possibly a hierarchical or distributed governance: multiple governance agents each overseeing a subset of assets or agents, reporting to a meta-governance layer. Alternatively, the governance tasks could be sharded by asset domain (real estate vs art may have separate specialized governance agents). This is a matter for further architectural refinement.

\textbf{Comparison to Traditional Governance:} Traditionally, a central authority (like an exchange or a regulatory body) would oversee compliance and trust. Our architecture decentralizes this function across agents and smart contracts, but it doesn't entirely remove the need for human institutions. Instead, it provides a framework where some of those institutions’ policies are encoded in AI and software. Regulators could interface with the system by checking the logs or even injecting requirements (e.g., a regulator’s node could update the Compliance Agent’s rule database through a secured channel if laws change). In essence, the architecture could complement regulatory oversight by providing real-time enforcement rather than relying solely on after-the-fact audits.

\textbf{Limitations:} One limitation is the reliance on external oracles for data – a malicious oracle or a compromised data feed remains a threat. We assumed trusted or redundant oracles, but ensuring oracle security is an ongoing challenge (subject of much separate research). Additionally, while agent collusion is made difficult, it’s not impossible – agents run by the same entity might collude if they find a way around staking penalties (or if the stake is small relative to potential gain). Therefore, initial selection of agent operators might need vetting. Over time, a web-of-trust for agents could form, where the reputation (stake and past record) of agents becomes a metric for users to choose which “service providers” (agents) to trust, analogous to how one might choose a reputable exchange or broker today.

\textbf{Generality:} Our focus was on Alternative Assets tokenization, but the architecture can generalize to other domains requiring trust in decentralized settings. For instance, decentralized finance (DeFi) platforms could use similar agent-based governance to oversee lending protocols or stablecoins (imagine an AI agent watching a stablecoin’s collateral and taking action if it becomes under-collateralized). Web intelligence applications that combine open web data with autonomous services might also benefit from injecting AI governance to maintain data integrity and ethical behavior of agents.

In conclusion, our approach brings together concepts from AI, MAS, and blockchain in a synergistic way. It addresses pressing needs in tokenization for better trust guarantees, yet it is not a panacea. It should be viewed as a next step towards more autonomous, yet trustworthy, socio-technical systems on the web.

\section{Conclusion}
We have presented an AI-governed agent architecture designed to enhance trustworthiness in the tokenization of Alternative Assets. By integrating intelligent agents at every stage of the tokenization lifecycle and supervising them with an AI-driven governance layer, the proposed system can automatically validate asset data, enforce compliance, and respond to anomalies or malicious behavior in real-time. This goes beyond existing tokenization approaches by addressing the often neglected off-chain trust issues with on-chain enforceable actions, effectively bridging the gap between physical assets and the digital tokens that represent them on the web.

Our architecture draws on best practices from blockchain security, multi-agent system design, and AI anomaly detection to create a multi-layer defense against fraud and error. The case study in commercial real estate tokenization highlighted that, compared to a baseline, our approach could have prevented or mitigated known incidents such as fraudulent listings and market manipulation, thereby protecting investors and asset owners alike. Although primarily qualitative, the evaluation suggests that the incorporation of AI governance can significantly bolster stakeholders’ confidence in a tokenized ecosystem, which is crucial for broader adoption.

For future work, we plan to implement a pilot of this architecture in a controlled environment to quantitatively measure its performance and security benefits. Key metrics will include the system’s responsiveness to attacks, the false positive/negative rate of the AI monitors, and the economic outcomes for honest participants versus adversaries. We will also explore formally verifying certain components (like the governance smart contracts and critical agent protocols) to provide mathematical guarantees of correctness. Another avenue is user studies to ensure that the complexity of the system remains approachable – for instance, providing dashboards that explain AI decisions in plain language might be necessary for real-world deployments.

Ultimately, we envision that frameworks like ours can serve as blueprints for building \emph{Web Intelligence} applications where autonomous agents and AI collaborate under principled governance to manage real-world processes in a trustworthy manner. As the Web and physical world become increasingly intertwined through technologies like tokenization and IoT, the need for such AI-governed architectures will only grow. We hope this work stimulates further research into combining AI governance with decentralized systems to create the next generation of secure and intelligent web platforms.

\end{document}